\documentclass[12pt]{article}
\usepackage[dvips]{color}
\usepackage{epsfig}
\usepackage{amsmath}
\usepackage{cite}
\usepackage{graphicx}
\usepackage{color}
\usepackage{subfigure}
\def\Box{\hbox{$\rlap{$\sqcup$}\sqcap$}}
\def\Box{\hbox{$\rlap{$\sqcup$}\sqcap$}}
\begin{document}
\begin{center}
\Large{\bf Further Refining Swampland dS Conjecture in \\ Mimetic f(G) Gravity}\\
\small \vspace{1cm}{\bf S. Noori Gashti$^{\star}$\footnote {Email:~~~saeed.noorigashti@stu.umz.ac.ir}}, \quad
{\bf J. Sadeghi$^{\star}$\footnote {Email:~~~pouriya@ipm.ir}}, \quad
and {\bf M. R. Alipour$^{\star}$\footnote {Email:~~~mr.alipour@stu.umz.ac.ir}}, \quad
\\
\vspace{0.5cm}$^{\star}${Department of Physics, Faculty of Basic
Sciences,\\
University of Mazandaran
P. O. Box 47416-95447, Babolsar, Iran}\\
\small \vspace{1cm}
\end{center}
\begin{abstract}
Mimetic gravity analysis has been studied as a theory in various types of general relativity extensions, such as mimetic f(R) gravity,  mimetic f(R, T) gravity, mimetic f(R, G) gravity, etc., in the literature. This paper presents a set of equations arising from mimetic conditions and studies cosmic inflation with a combination of mimetic f(G) gravity and swampland dS conjectures. We analyze and evaluate these results. Therefore, we first thoroughly introduce the mimetic f(G) gravity and calculate some cosmological parameters such as the scalar spectral index, the tensor-to-scalar ratio, and the slow-roll parameters. Also, we investigate the potential according to the mimetic f(G) gravity. Then we will challenge the swampland dS conjectures with this condition. By expressing the coefficient of swampland dS conjectures viz $C_{1}$ and $C_2$ in terms of $n_{s}$ and $r$, we plot some figures and determine the allowable range for each of these cosmological parameters and these coefficients, and finally, compare these results with observable data such as Planck and BICEP2/Keck array data. We show $C_{1}$ and $C_2$ are not  $\mathcal{O}(1)$, so the refining swampland dS conjecture is not satisfied for this inflationary model. Then we examine it with further refining swampland dS conjecture, which has a series of free parameters such as $a,b>0$, $q>2$, and $a+b=1$. By adjusting these parameters, the compatibility of the mentioned conjecture with the inflationary model can be discussed. We determine the further refining swampland dS conjecture is satisfied. when $a < \frac{1}{1.00489}=0.99513$, we can always find $a$, $b$ and $q$ whose value is larger than 2,
viz for $q=2.4$, we find $0.99185\leq a < 1$, which we can choose $a=0.99235$ according to the condition $a < 0.99513$. Also we know $b=1-a$, so we will have $1-0.99235=0.00765 > 0$. 
\end{abstract}
\newpage
\tableofcontents
\section{Introduction}
The introduced inflation has been extensively investigated concerning conditions such as slow-roll, fast-roll, constant-roll, etc. It has explicitly occurred as a period in the early stages of universe evolution with an early accelerated expansion. In a small moment, a microscopic universe turned into a macroscopic one and responded well to many of the Big Bang problems that the universe's initial conditions. Besides, inflation does not create a symmetric universe since quantum fluctuations grow rapidly and become macroscopic perturbations. These quantum fluctuations have provided the primary seeds for forming such a universe structure, and all of these are the major success of the inflation scenario. Of course, alternative strategies such as the ekpyrotic scenario have been used, but they all have a super-accelerating phase\cite{1,1a,2,2a,3,3a}. So far, different types of inflation models have been studied. Still, over the past decade, with the missions that the probe  (WMAP)\cite{4,4a}, and more recently Planck mission and BICEP2 \cite{5,5a,5b,5c,5d}, a way to measure the power spectrum of primordial perturbations have been identified, there have been created many differences between the inflation models.
Some of these models have more credibility and value. It explains inflation from its slow roll condition by introducing a scalar field that mimics a cosmic constant during inflation. At the end of inflation, it rolls down the step of its potential despite the assumption of the emergence of different particles from decaying the scalar field. It leads to reheating of the universe, which is associated with the initial state of the Big Bang\cite{1,6,6a}. In addition to conventional scalar field models, several generalized theories of general relativity are used to study inflation, including the effective theories of gravity as f(R),  f(R, T), etc.
Other cosmological structures, such as dark energy and dark matter, are also studied with these models.
One of the most important models is the so-called $R^{2}$ inflation or the Starobinsky model.
Other extensions of general relativity are also used in cosmological studies, such as Gauss-Bonnet gravities, created by inserting a nonlinear f(G) function into action. Besides, structures such as mimetic f(R) gravity, mimetic f(R, T) gravity, and mimetic f(R, G) gravity have been studied to study inflation and other cosmological applications. These results have been compared with recent observable data. The cosmological implications of each of these models have been thoroughly studied. Nonlinear Gauss-Bonnet terms in action can also cause a series of instabilities in the anisotropic universe as Kasner-type background. Still, such instabilities are not seen (FLRW) cosmic structure \cite{7,7a,7b,7c,7d,7e,7f,8,9,10,11,11a,11b,12,13,13a,13b,13c,13d,14,14a,14b,14c,14d,14e,14f,15,15a,16,16a,16b,16c,16d,16e,16f,17,18}. There are several observational reasons that somehow prove that, in addition to the dark energy content, most matter in our universe is made up of an unknown fluid that somehow has a gravitational interaction and behaves like a pressureless fluid named dark matter. Although there are many descriptions of dark matter, its nature is still unclear. Some models describe it as a particle. Some models also use unique mechanics to describe the geometry of dark matter, known as mimetic gravity, which leads to behavior similar to that of a pressureless fluid with changes in action. Different types of these models have been introduced in the literature, and each offers a solution to dark matter and dark energy. These models fully explain inflation and, early universe dynamics, the growth of cosmic perturbations.\cite{19,20,20a,20b,20c,20d,21,22,23,24,25,26,27,28,29,30,31,32,33,34}. Other classical aspects of mimetic gravity have also been used as symmetrical spherical solutions as black holes\cite{24,35}. Given all the above, in this article, we want to evaluate a particular type of cosmological implications of mimetic gravity that has not been studied in the literature so far. According to the monitoring of swampland dS conjectures, that is the study of inflation from the mimetic gravity point of view. We challenge combining these ideas, calculate cosmological parameters and other cosmological quantities, compare the results with observable data, and analyze the compatibility or incompatibility of this model with swampland dS conjectures. The swampland program has recently been studied in many cosmological studies: inflation, black hole physics, string theory, dark matter, dark energy, and many other cosmic theories. Its various cosmological applications have been analyzed. Recently, new conjectures have been added to this spectrum, and the relationship between these conjectures has been thoroughly investigated. In some cases, the existing flaws in these conjectures have been eliminated with corrections. So they found them somehow in line with observable data.
Basically, in the swampland program, concepts such as swampland include theories consistent with string theory and incompatible with quantum gravity, and in contrast to the landscape, which is somehow surrounded by the swampland, and include theories that are consistent with quantum gravity. This article will use the refined swampland dS conjecture that we will explain in detail in the following sections. For more on swampland dS conjecture and its cosmological applications can be found in Ref.s \cite{36,37,38,39,39a,39b,40,40a,41,41a,41b,42,42a,43,43a,44,45,46,47,48,49,50,51,52,53,54,55,56,56a,56b,57,58,59,60,61,61ab,62ab,62ac}. So this paper is organized as follows.\\
In section 2, we overview the swampland dS conjecture. In sections 3 and 4, we introduce the mimetic $f(G)$ gravity theory, inflation's aspects in mimetic Gauss-Bonnet gravity, and the extensions of mimetic f(G) gravity. In section 5, we first calculate some cosmological parameters, such as the scalar spectral index, the tensor-to-scalar ratio, and the slow-roll parameters. Also, we investigate the potential according to the mimetic f(G) gravity. Then we will challenge the swampland dS conjectures with this condition. By expressing the coefficient of swampland dS conjectures in terms of $n_{s}$ and $r$, we plot some figures and determine  the allowable range for each of these cosmological parameters and these coefficients, and finally, compare these results with observable data such as Planck and BICEP2/Keck array data, and second, we challenging the model with recent further refining swampland dS conjecture and discuss the result; finally, we conclude in section 6

\section{Swampland dS conjectures}
Recently, a set of conjectures in the form of a swampland program has been introduced to investigate the cosmological implications of string theory. String theory has made significant advances in recent years to introduce solutions to quantum gravity, but it has struggled in presenting inflation scenarios.
It can be briefly stated that the creation of dS solutions in this theory has faced severe challenges.
Even if several possible routes are suggested\cite{a}, of course, this is in contrast to the supergravity situation $N=1$, in which successful inflation models are obtained in such a framework\cite{b}.
This is strange to some extent because it is thanks to inflation that specific issues of superstring models with intermediate-scale can be bypassed \cite{c}.
Of course, in addition to single-field inflation, other models, such as two-field inflation, have been presented, which, of course, have not been systematically studied in string theory\cite{d}.
The current conjectures that some dS solutions in string theory can not be put into landscape leads to a group of consistent-looking theories that do not admit a suitable ultraviolet completion in string theory and fall into the swampland. The big question is whether the resulting conditions are compatible with inflation.
This collection of conjectures contains a series of general criteria. The considerable importance is that charge symmetries are assumed to be local gauge symmetries so that at least one particle must have a mass in Planck units more minor than the gauge coupling strength to confirm that gravity is weak\cite{e,e1}.
Of course, the question remains whether string theory can accept a de Sitter vacuum.
There is no complete top-down de Sitter structure in a controllable string theory regime \cite{45,46,47,48,49,50,51,52,53,54,55}.
This has become a challenging goal in string theory because of the apparent cosmological implications of our universe.
Therefore, since string theory also describes the structure of quantum gravity, these conjectures derived from string theory are also examined to study its cosmological consequences and consistency with QG.
As we have said, these conjectures have a set of criteria based on the physical explanation of these conjectures; mathematics is introduced to execute these conjectures. Different types of theories can be tested with these conjectures \cite{45,46,47,48,49,50,51,52,53,54,55}.
Those theories consistent with these criteria are put in landscape and adapted to quantum gravity, which can be a fascinating result and a new way of exploring as much as possible to find a solution to quantum gravity. To study on a large scale or even soon to be able to see the parameters or structures of it \cite{45,46,47,48,49,50,51,52,53,54,55}.
Hence, the above discussions motivated us to challenge effective theories such as f(G) gravity with refined swampland dS conjecture and examine its cosmological consequences.
Examining different theories from this perspective can help identify ideas compatible with quantum gravity and even a new classification for inflation models.
The primary purpose of this section is to challenge three combinations: inflation, mimetic f(G) gravity, and swampland dS conjectures. These theories have been studied extensively in the literature, and their cosmological applications have been reviewed. The results of these studies have been compared with observable data. But in this article, we examine a specific type of these applications and analyze the results. Therefore, by introducing our inflation model and combining it with mimetic f(G) gravity and swampland dS conjectures, we calculate a series of cosmological parameters such as scalar spectral index and tensor-to-scalar ratio, potential, etc. We determine the exact values of these cosmological parameters and swampland dS conjecture coefficients by combining these cosmological parameters arising from mimetic f(G) gravity with swampland dS conjectures. We plot figures to determine the allowable range of each parameter. We will analyze the results and compare them with the observable data\cite{5} and investigate the compatibility or incompatibility of the swampland dS conjectures with this mimetic f(G) gravity. The swampland program, mainly the  swampland dS conjecture, has been studied in detail in many cosmological subjects, such as inflation, the physics of black holes and dark energy, and their cosmic applications. Recently, new conjectures have been added to the swampland program. There has been a lot of discussion about the relationship among these conjectures\cite {45,46,47,48,49,50,51,52,53,54,55}. But in this article, we use the refined swampland dS conjecture expressed in the following form,
\begin{equation*}\label{399}
|\nabla V|\geq\frac{C_{1}}{M_{p}}V, \hspace{12pt}  min \nabla\partial V\leq -\frac{C_{2}}{M_{pl}^{2}}V.
\end{equation*}
The above equations for the $V>0$, which is given by,
\begin{equation*}\label{577}
\sqrt{2\epsilon_{V}}\geq C_{1} ,\hspace{12pt}  or\hspace{12pt} \eta_{V}\leq -C_{2},
\end{equation*}
where $C_{1}$ and $C_{2}$ are constant parameters. With all the above descriptions, we want to consider a simple model and thoroughly study the above concepts: the model that describes cosmological evolution using the Hubble parameter.
\section{mimetic $f(G)$ gravity}
In this section, we introduce mimetic f (G) gravity. Mimetic gravity with a physical metric in terms of an auxiliary metric and scalar field is the following form for challenging the conformal degree of freedom\cite{21},
\begin{equation}\label{1}
g_{\mu\nu}=-\widehat{g}^{\rho\sigma}\partial_{\rho}\phi\partial_{\sigma}\phi\widehat{g}_{\mu\nu}.
\end{equation}
The physical metric is invariant under the conformal transformation $\widehat{g}_{\mu\nu}\rightarrow\Omega^{2}g_{\mu\nu}$, therefore according to equation (1), we will have the following constraint for the scalar field,
\begin{equation}\label{2}
g^{\mu\nu}\partial_{\mu}\phi\partial_{\nu}\phi=-1.
\end{equation}
After the above explanation, we introduce the action of Einstein Hilbert in the theory of mimetic gravity with respect to physical metrics $g_{\mu\nu}$,
\begin{equation}\label{3}
S=\frac{1}{2k^{2}}\int d^{4}x\sqrt{-g}R(g_{\mu\nu}.
\end{equation}
The field equations by varying  the action with respect to the metric are the following form,
\begin{equation}\label{4}
\begin{split}
(G^{\mu\nu}-T^{\mu\nu})+(G-T)g^{\mu\lambda}g^{\mu\gamma}\partial_{\lambda}\phi\partial_{\gamma}\phi=0,\\
\nabla_{\mu}\{(G-T)\partial^{\mu}\phi\}=0.
\end{split}
\end{equation}
We describe the field equations exclusively in terms of the scalar field and  $g_{\mu\nu}$  and scalar field with respect to the conformal degree of freedom. The important point here is the mimetic field, which can be introduced and behave as a pressureless fluid and can be interpreted as a contribution to dark matter. It leads to appear a new dynamical degree of freedom. The action (3) is not unique and can also be extended to explain dark energy\cite{21,22,23,24,25}. With a non-linear function of the Gauss-Bonnet topological invariant, this paper extends the mimetic gravity and examines the inflation in this structure by challenging the swampland dS conjectures. So the action for f(G) gravity is expressed,
\begin{equation}\label{5}
S=\int d^{4}x\sqrt{-g}\Big(\frac{R}{2k^{2}}+f(\mathcal{G})\Big)+\mathcal{S}_{m},
\end{equation}
with respect to equation (5), the Gauss-Bonnet term $\mathcal{G}$ which is given by,
\begin{equation}\label{6}
\mathcal{G}=R^{2}-4R_{\mu\nu}R^{\mu\nu}+R_{\mu\nu\lambda\sigma}R^{\mu\nu\lambda\sigma},
\end{equation}
where $R_{\mu\nu}$ and $R_{\mu\nu\lambda\sigma}$ are the Ricci tensor and Riemann tensor, respectively. Also, we assume $k^{2}=1$. After the above explanations, we will rewrite the field equations following the mimetic gravitational theory mentioned in this paper. Therefore, according to the metric $g_{\mu\nu}$ and the auxiliary field in equation (1), the field equations are expressed in the following by considering the auxiliary metric $\widehat{g}_{\mu\nu}$ as the main mimetic example, and the action (3)\cite{17,21,22,23,24,25,26,27},
\begin{equation}\label{7}
\begin{split}
&R_{\mu\nu}-\frac{1}{2}Rg_{\mu\nu}+\Big(f_{\mathcal{G}}\mathcal{G}-f(\mathcal{G})\Big)g_{\mu\nu}+\\
&8\Big\{R_{\mu\rho\nu\sigma}+R_{\rho\nu}g_{\sigma\mu}-R_{\rho\sigma}g_{\nu\mu}-R_{\mu\nu}g_{\sigma\rho}+R_{\mu\sigma}g_{\nu\rho}+\frac{R}{2}(g_{\mu\nu}g_{\sigma\rho}-g_{\mu\sigma}g_{\nu\rho})\Big\}\nabla^{\rho}\nabla^{\sigma}f_{\mathcal{G}}\\
&+\partial_{\mu}\phi\partial_{\nu}\phi\Big(-R+8(-R_{\rho\sigma}+\frac{1}{2}Rg_{\rho\sigma})\nabla^{\rho}\nabla^{\sigma}f_{\mathcal{G}}+4(f_{\mathcal{G}}\mathcal{G}-f(\mathcal{G}))\Big)=T_{\mu\nu}+\partial_{\mu}\phi\partial_{\nu}\phi T,\\
\end{split}
\end{equation}
\begin{equation}\label{8}
\nabla^{\mu}\Big(\partial_{\mu}\phi(-R+8(-R_{\rho\sigma}+\frac{1}{2}Rg_{\rho\sigma})\nabla^{\rho}\nabla^{\sigma}f_{\mathcal{G}}+4(f_{\mathcal{G}}\mathcal{G}-f(\mathcal{G}))-T)\Big)=0.
\end{equation}
After introducing the field equations, we proceed to our studies in flat FRW space-times so we will have,
\begin{equation}\label{9}
ds^{2}=-dt^{2}+a(t)^{2}\delta_{ij}dx^{i}dx^{j},
\end{equation}
where $a$ is the scale factor. Also the scalar curvature $R$ and e Gauss-Bonnet term $\mathcal{G}$ which are following form,
\begin{equation}\label{10}
R=6(\frac{\ddot{a}}{a}+\frac{\dot{a}^{2}}{a^{2}})=6(\dot{H}+2H^{2}),
\end{equation}
and
\begin{equation}\label{11}
\mathcal{G}=-24\frac{\ddot{a}\dot{a}^{2}}{a^{3}}=-24H^{2}(\dot{H}+H^{2}).
\end{equation}
So, with respect to equation (7) and FLRW spacetime, we can obtain,
\begin{equation}\label{12}
8H^{2}\frac{\partial^{2}f_{\mathcal{G}}}{\partial t^{2}}+16H(\dot{H}+H^{2})\frac{\partial f_{\mathcal{G}}}{\partial t}-(f_{\mathcal{G}}\mathcal{G}-f)+2\dot{H}+3H^{2}=-p.
\end{equation}
Also, considering equation (8), we have the following relation,
\begin{equation}\label{13}
4H^{2}\frac{\partial^{2}f_{\mathcal{G}}}{\partial t^{2}}+4H(2\dot{H}+3H^{2})\frac{\partial f_{\mathcal{G}}}{\partial t}+\frac{2}{3}(f_{\mathcal{G}}\mathcal{G}-f)+\dot{H}+2H^{2}=-\frac{c}{a^{3}}-\frac{\rho}{6}+\frac{p}{2},
\end{equation}
where $c$ is a constant integration parameter, if we consider a relationship to form $T_{\mu\nu}=(p+\rho)u_{\mu}u_{\nu}+pg_{\mu\nu}$ for a perfect fluid\cite{17,21,22,23,24,25,26,27}, the right side of equation (13) is obtained after the integration of equation (8), which in a way represents the behavior of a pressureless perfect fluid or, as we mentioned in the text, represents mimetic dark matter. Hence, we will have to combine two equations, (12) and (13). So,
\begin{equation}\label{14}
4H^{2}\frac{dg(t)}{dt}+4H(2\dot{H}-H^{2})g(t)=B(t)=-\dot{H}-\frac{1}{2}(p+\rho)-\frac{c}{a^{3}},
\end{equation}
where $g(t)=\frac{df_{\mathcal{G}}}{dt}$. The important point here is that for a particular ansatz for the Hubble parameter, equation (14) is solved for g(t) and will be reconstructed mimetic gauss-bonnet action using equations (10) and (11). For example, assuming that the universe is filled only with dust and mimetic dark matter, we will have $p=0$ and $\rho=\rho_{0}a(t)^{-3}$. So $B=-\dot{H}-\widehat{c}a(t)^{-3}$ and $\widehat{(c)}=c+\frac{\rho_{0}}{2}$. Hence the general form for g(t) will be expressed in the following form,
\begin{equation}\label{15}
g(t)=\Big(\frac{H_{0}}{H(t)}\Big)^{2}\exp\Big(\int_{0}^{t}H(t_{1})dt_{1}\Big)\Big\{g_{0}+\frac{1}{4H_{0}^{2}}\int_{0}^{t}\exp\Big(-\int_{0}^{t_{2}}H(t_{1})dt_{1}\Big)B(t_{2})dt_{2}\Big\}.
\end{equation}
Therefore, for a suitable Hubble parameter,  action is calculated. In the following, by introducing the inflation model from the point of view of mimetic gauss-bonnet gravity and the extensions of mimetic f(G) gravity, we challenge a particular type of inflation model according to this gravitational model and the swampland dS conjectures.
\section{Inflation \& mimetic f(G) gravity}
In this section, we examine inflation from the point of view of mimetic Gauss-Bonnet gravities and introduce some cosmological parameters. Also, in the continuation of this section, we will discuss the expanded structure of f(G) mimetic gravity. Now, with respect to the action, equation (5), we will have,\cite{17,21,22,23,24,25,26,27}
\begin{equation}\label{16}
S_{\phi}=\int d^{4}x\sqrt{-g}\Big\{-\frac{1}{2}\partial_{\mu}\phi\partial^{\mu}\phi-V(\phi)\Big\}.
\end{equation}
So, considering equation (16), the FRW equations which are given by,
\begin{equation}\label{17}
\begin{split}
&\frac{3}{k^{2}}H^{2}=\frac{1}{2}\dot{\phi}^{2}+V(\phi),\\
&-\frac{1}{k^{2}}(3H^{2}+2\dot{H})=\frac{1}{2}\dot{\phi}^{2}-V(\phi),
\end{split}
\end{equation}
also,
\begin{equation}\label{18}
\begin{split}
\ddot{\phi}+3H\dot{\phi}+\frac{\partial V(\phi)}{\partial\phi}=0.
\end{split}
\end{equation}
In general, we consider several fields as $N=\ln\Big(\frac{a(t)}{a(0)}\Big)$ as an independent variable. the equation (17) with respect to $N=\phi$ are expressed in the following form,
\begin{equation}\label{19}
\begin{split}
&\omega(\phi)=-\frac{2H'(\phi)}{k^{2}H(\phi)},\\
&V(\phi)=\frac{1}{k^{2}}\Big(3(H(\phi))^{2}+H(\phi)H'(\phi)\Big),
\end{split}
\end{equation}
here $\omega(\phi)$ represents the kinetic term from redefining the scalar field. We will now define the slow-roll parameters according to the concepts mentioned earlier in the following form according to slow-roll inflation,
\begin{equation}\label{20}
\begin{split}
\epsilon=\frac{1}{2k^{2}}\Big(\frac{V'(\phi)}{V(\phi)}\Big)^{2},\hspace{12pt}\eta=\frac{1}{k^{2}}\frac{V''(\phi)}{V(\phi)},\hspace{12pt}\lambda^{2}=\frac{1}{k^{4}}\frac{V'(\phi)V'''(\phi)}{\Big(V(\phi)\Big)^{2}}.
\end{split}
\end{equation}
We introduce the scalar spectral index $n_{s}$ and tensor-to-scalar ratio $r$ in terms of the slow parameter as follows,
\begin{equation}\label{21}
\begin{split}
n_{s}=1-6\epsilon+2\eta,\hspace{12pt}r=16\epsilon,\hspace{12pt}\alpha_{s}=\frac{dn_{s}}{d\log k}\approx 16\epsilon\eta-24\epsilon^{2}-2\xi^{2}.
\end{split}
\end{equation}
By using the equation (19), the slow parameter rewrite in terms of Hubble parameter H \cite{17,21,22,23,24,25,26,27} which are obtained as,
\begin{equation}\label{22}
\begin{split}
\epsilon=-\frac{H(N)}{4H'(N)}\Big[\frac{H''(N)H(N)+6H'(N)H(N)+H'^{2}(N)}{3H^{2}(N)+H'(N)H(N)}\Big]^{2},
\end{split}
\end{equation}
\begin{equation}\label{23}
\begin{split}
\eta=-\frac{\Big(9\frac{H'(N)}{H(N)}+3\frac{H''(N)}{H(N)}+\frac{1}{2}(\frac{H'(N)}{H(N)})^{2}+3\frac{H''(N)}{H'(N)}+\frac{H'''(N)}{H'(N)}-\frac{1}{2}(\frac{H''(N)}{H'(N)})^{2}\Big)}{2\Big(3+\frac{H'(N)}{H(N)}\Big)},
\end{split}
\end{equation}
prime is a derivative of N. According to slow-roll parameters in equations (22) and (23), the values of cosmic parameters such as scalar spectral index $n_{s}$ and tensor-to-scalar ratio $r$ can be calculated for different types of inflation models. The values of these parameters with respect to recent observable data\cite{5} are expressed in the following form,
\begin{equation}\label{24}
\begin{split}
n_{s}=0.968\pm0.006, \hspace{12pt}r<0.07.
\end{split}
\end{equation}
Of course, in previous missions, more accurate values for these parameters have been obtained as $n_{s}=0.9644\pm0.0049$ and $r<0.10$, and in BICEP2/Keck Array, the tensor-to-scalar ratio is as $r<0.07$\cite{5}. In the continuation of this article, we introduce the extensions of mimetic f(G) gravity. Finally, we mention our inflation model and call the swampland dS conjectures. Then we examine new challenges and analyze the results. Hence, another way to explain the mimetic gravity is to use a special constraint added to the action through a Lagrange multiplier, expressed in the following form,
\begin{equation}\label{25}
\begin{split}
S=\int d^{4}x\sqrt{-g}\Big\{\frac{R}{2k^{2}}+f(\mathcal{G})+\lambda(g^{\mu\nu}\partial_{\mu}\phi\partial_{\nu}\phi+1)\Big\}.
\end{split}
\end{equation}
Field equivalents can be recovered for the mimetic f(G) gravity equation (7). In this section, we want to examine a series of extensions for the above action whose dynamic form by adding the kinetic sentence of a potential to the scalar field \cite{17,21,22,23,24,25,26,27}. Hence we will have,
\begin{equation}\label{26}
\begin{split}
S=\int d^{4}x\sqrt{-g}\Big\{\frac{R}{2k^{2}}+f(\mathcal{G})-\varepsilon g^{\mu\nu}\partial_{\mu}\phi\partial_{\nu}\phi-V(\phi)+\lambda(g^{\mu\nu}\partial_{\mu}\phi\partial_{\nu}\phi+1)\Big\}.
\end{split}
\end{equation}
So with respect to equations (9), (26), and metric $g_{\mu\nu}$, the equations of motion are calculated as,
\begin{equation}\label{27}
\begin{split}
3H^{2}+24H^{3}\frac{df_{\mathcal{G}}(\mathcal{G})}{dt}+f(\mathcal{G})-f_{\mathcal{G}}(\mathcal{G})\mathcal{G}=\varepsilon\dot{\phi}^{2}+V(\phi)-\lambda(\dot{\phi}^{2}+1),
\end{split}
\end{equation}

\begin{equation}\label{28}
\begin{split}
-2\dot{H}-3H^{2}-8H^{2}\frac{d^{2}f_{\mathcal{G}}(\mathcal{G})}{dt^{2}}-16H(\dot{H}+H^{2})\frac{df_{\mathcal{G}}(\mathcal{G})}{dt}-f(\mathcal{G})+f_{\mathcal{G}}(\mathcal{G})\mathcal{G}=\varepsilon\dot{\phi}^{2}-V(\phi)-\lambda(\dot{\phi}^{2}-1).
\end{split}
\end{equation}
The restriction for the scalar field $\phi$ with respect to variation about Lagrange multiplier is given by: $\phi$: $$\dot{\phi}^{2}=1.$$
So by combining the two equations (27) and (28) and redefining the scalar field as $\phi=t$, the following expressions for potential and Lagrange multiplier can be obtained \cite{17,21,22,23,24,25,26,27},
\begin{equation}\label{29}
\begin{split}
V(t)=2\dot{H}+3H^{2}+\varepsilon\dot{\phi}^{2}+8H^{2}\frac{d^{2}f_{\mathcal{G}}(\mathcal{G})}{dt^{2}}+16H(\dot{H}+H^{2})\frac{df_{\mathcal{G}}(\mathcal{G})}{dt}+f(\mathcal{G})-f_{\mathcal{G}}(\mathcal{G})\mathcal{G},
\end{split}
\end{equation}

\begin{equation}\label{30}
\begin{split}
\lambda(t)=\dot{\phi}^{-2}\Big(\frac{\rho}{2}+\dot{H}+4H(2\dot{H}-H^{2})\frac{df_{\mathcal{G}}(\mathcal{G})}{dt}+4H^{2}\frac{d^{2}f_{\mathcal{G}}(\mathcal{G})}{dt^{2}}\Big)+\varepsilon.
\end{split}
\end{equation}
Therefore, the action (26) can be easily reconstructed for each inflation model according to the specified value of the Hubble parameter. Of course, the potential and  Lagrange multiplier can also be expressed in terms of the number of e-folds $N$. Hence the two equations (29) and (30) are rewritten as follows,
\begin{equation}\label{31}
\begin{split}
V(N)=&2H(N)H'(N)+3H^{2}(N)+\varepsilon H^{2}(N)\dot{\phi}^{2}+8H^{2}(N)\Big\{H^{2}(N)\frac{d^{2}}{dN^{2}}+H(N)H'(N)\frac{d}{dN}\Big\}\\
&+16H^{3}(N)\Big\{H'(N)+H(N)\Big\}\frac{df_{\mathcal{G}}(\mathcal{G})}{dN}-f_{\mathcal{G}}(\mathcal{G})\mathcal{G}+f(\mathcal{G}),
\end{split}
\end{equation}

\begin{equation}\label{32}
\begin{split}
\lambda(N)=&\frac{\rho}{2}+4H(N)H'(N)+H^{3}(N)\Big\{2H'(N)-H(N)\Big\}\frac{df_{\mathcal{G}}(\mathcal{G})}{dN}\\
&+4H^{2}(N)\Big\{H^{2}(N)\frac{d^{2}}{dN^{2}}+H(N)H'(N)\frac{d}{dN}\Big\}f_{\mathcal{G}}(\mathcal{G})+\varepsilon.
\end{split}
\end{equation}
We can define the function $f(\mathcal{G})$ viz equation (26) in terms of an extra auxiliary field. So, the new action is written as follows \cite{17,21,22,23,24,25,26,27},
\begin{equation}\label{33}
\begin{split}
S=\int d^{4}x\sqrt{-g}\Big\{\frac{R}{2k^{2}}+\varphi\mathcal{G}-U(\varphi)-V(\phi)+\lambda(g^{\mu\nu}\partial_{\mu}\phi\partial_{\nu}\phi+1)\Big\}.
\end{split}
\end{equation}
where $\varphi=f'(\mathcal{G})$ and we have omitted the kinetic term for $\phi$ for simplicity.  Now the variation of the above action according to $g_{\mu\nu}$ will be expressed in the following form,
\begin{equation}\label{34}
\begin{split}
R_{\mu\nu}-\frac{1}{2}Rg_{\mu\nu}+U(\varphi)g_{\mu\nu}+2H_{\mu\nu}-\Big\{\lambda(g^{\mu\nu}\partial_{\mu}\phi\partial_{\nu}\phi+1)-V(\phi)\Big\}g_{\mu\nu}+2\lambda\partial_{\mu}\phi\partial_{\nu}\phi=0,
\end{split}
\end{equation}
here the term $H_{\mu\nu}$ is the variation of $\varphi\mathcal{G}$,
\begin{equation}\label{35}
\begin{split}
H_{\mu\nu}\equiv &\frac{1}{\sqrt{-g}}\frac{\delta(\sqrt{-g}\varphi\mathcal{G})}{\delta g^{\mu\nu}}=2R(g_{\mu\nu}\Box-\nabla_{\mu}\nabla_{\nu})\varphi+4R^{\alpha}_{\mu}\nabla_{\alpha}\nabla_{\nu\varphi}+4R^{\alpha}_{\nu}\nabla_{\alpha}\nabla_{\mu\varphi}\\
&-4R_{\mu\nu}\Box\varphi-4g_{\mu\nu}R^{\alpha\beta}\nabla_{\alpha}\nabla_{\beta\varphi}+4R_{\alpha\mu\beta\nu}\nabla^{\alpha}\nabla^{\beta}\varphi.
\end{split}
\end{equation}
Now, we can express the field equations\cite{17,21,22,23,24,25,26,27} according to equation (34),
\begin{equation}\label{36}
\begin{split}
3H^{2}(t)\Big(1+8H(t)\varphi'(t)\Big)+\lambda(t)\Big(1+\phi'^{2}(t)\Big)-V\Big(\phi(t)\Big)-U\Big(\varphi(t)\Big)=0,
\end{split}
\end{equation}

\begin{equation}\label{37}
\begin{split}
&H(t)\Big\{16H^{2}(t)\varphi'(t)+16\dot{H}(t)\varphi'(t)+H(t)\Big(3+\varphi''(t)\Big)\Big\}\\
&+\lambda(t)\Big\{1-\phi'^{2}(t)\Big\}+2\dot{H}(t)-V\Big(\phi(t)\Big)-U\Big(\varphi(t)\Big)=0.
\end{split}
\end{equation}
Considering all the concepts mentioned above, in the continuation of this article, we examine the cosmological implications of selecting a suitable form of an inflation model and combining different conditions, such as mimetic f(G) gravity and swampland dS conjectures from this combination. We calculate each of these cosmological parameters, such as the scalar spectral index and the tensor-to-scalar ratio, slow-roll parameters, potential, and Lagrange multiplier, and by combining these parameters with swampland dS conjectures, we challenge the compatibility of these models with these conditions according to the observable data\cite{5}. Also, we determine the exact value of these parameters and components of the swampland dS conjectures and compare them with observable data\cite{5}. Therefore, we select a specific ansatz for the function f(G) to continue working in the below form. Here, we suppose a special ansatz with the most explicit probable selection for the f(G),
\begin{equation}\label{eq377}
f(\mathcal{G})=A\mathcal{G}^{n}
\end{equation}
where $A$ and $n$ are the constant parameters.
\section{Discussion and result}
In the present studies, different models can be challenged according to the concepts mentioned in this article, and the results can be explained entirely. Since this issue is analyzed for the first time, we intend to fully explain all aspects by choosing such a model. We study all the details and review cosmological applications and the conformity of the model with the observable data. According to the above description, we consider an inflationary model, which is described with the Hubble rate, to investigate cosmological evolution. i.e.,
\begin{equation}\label{eq388}
H(N)=(G_{0}+NG_{1})^{\theta}
\end{equation}
where $G_{0}$, and $G_{1} $ are constant parameter and $N$ is the number of e-folds. we consider the cases where the following Hubble parameter describes the cosmological evolution.
The two constant parameters are determined as $G_{1}<0$, $G_{0} > 0$. We have $|\dot{H}|\ll H^{2} $ during the inflationary period, which causes $\frac{G_{0}}{G_{1}}\gg N$, such that the Hubble parameter is roughly constant (de Sitter) during the inflation. After the end of inflation, the second sentence evolves vital, and the Hubble rate decays.
The purpose is studies of slow-roll inflation, and its concepts are fully described. So, according to equations (22), (23) and \eqref{eq388}, we will have,
\begin{equation}\label{eq38}
\epsilon=-\frac{G_{1}\theta\Big(6G_{0}+G_{1}(-1+6N+2\theta)\Big)^{2}}{4(G_{0}+G_{1}N)\Big(3G_{0}+G_{1}(3N+\theta)\Big)^{2}},
\end{equation}

\begin{equation}\label{eq39}
\eta=\frac{G_{1}\Big(G_{0}(8-26\theta)+G_{1}\Big(1+N(8-26\theta)+4\theta-6\theta^{2}\Big)\Big)}{4(G_{0}+G_{1}N)\Big(3G_{0}+G_{1}(3N+\theta)\Big)}.
\end{equation}
Also, we calculated the tensor-to-scalar ratio according to the above equations and equation (21) in the following form,
\begin{equation}\label{eq40}
r=-16\frac{G_{1}\theta\Big(6G_{0}+G_{1}(-1+6N+2\theta)\Big)^{2}}{4(G_{0}+G_{1}N)\Big(3G_{0}+G_{1}(3N+\theta)\Big)^{2}},
\end{equation}

\begin{equation}\label{eq41}
\begin{split}
n_{s}-1=&2\frac{G_{1}\Big(G_{0}(8-26\theta)+G_{1}\Big(1+N(8-26\theta)+4\theta-6\theta^{2}\Big)\Big)}{4(G_{0}+G_{1}N)\Big(3G_{0}+G_{1}(3N+\theta)\Big)}\\
&+6\frac{G_{1}\theta\Big(6G_{0}+G_{1}(-1+6N+2\theta)\Big)^{2}}{4(G_{0}+G_{1}N)\Big(3G_{0}+G_{1}(3N+\theta)\Big)^{2}}.
\end{split}
\end{equation}
To compare this inflation model with observable data, we need to determine the values of a series of cosmological parameters. So, by considering the free parameters as $(G_{0}=0.75, G_{1}=-0.00699, \theta=0.2 \hspace{0.2cm} and \hspace{0.2cm}  N=60)$, the two cosmological parameters, i.e., the scalar spectral index and the tensor-to-scalar ratio, are obtained $(n_{s}=0.964569)$, and $(r=0.06836)$, respectively, which are consistent with Planck's observable data. By placing equation \eqref{eq40} in each of the equations (31) and (32) with respect to $\phi=t$, we obtain the potential and Lagrange multiplier for the scalar field in terms of (N), which are expressed in the following form,
\begin{equation}\label{eq42}
\begin{split}
&V(N)=\frac{1}{3(G_{0}+G_{1}N)(G_{0}+G_{1}N)^{2\theta}}\big(3G_{0}++3G_{1}N+2G_{1}\theta\big)-\frac{24^{n}A(-1+n)}{\big(G_{0}+G_{1}(N+\theta)\big)^{2}}\\
&\times3\big(G_{0}+G_{1}N\big)^{4}+G_{1}\big(G_{0}+G_{1}N\big)\bigg(9(G_{0}+G_{1}N)^{2}+2n\bigg[4G_{0}^{2}+G_{0}G_{1}(-3+8N)\\
&+G_{1}^{2}(1+N(-3+4N))\bigg]\bigg)\theta+G_{1}^{2}\bigg(n^{2}(G_{1}-4G_{0}-4G_{1}N)^{2}+9(G_{0}+G_{1}N)^{2}+n(G_{0}\\
&+G_{1}N)\big(-5G_{1}+12G_{0}+12G_{1}N\big)\theta^{2}+G_{1}^{3}\bigg(nG_{1}+3(G_{0}+G_{1}N)\\
&+8n^{2}\big(-G_{1}+4G_{0}+4G_{1}N\big)\bigg)\theta^{3}+4n(-1+4n)G_{1}^{4}\theta^{4}\bigg)\\
&\times\bigg(-(G_{0}+G_{1}N)^{-1+4\theta}\big(G_{0}+G_{1}(N+\theta)\big)\bigg)^{n}+\varepsilon,
\end{split}
\end{equation}
and
\begin{equation}\label{eq43}
\begin{split}
&\lambda(N)=\frac{1}{24(G_{0}+G_{1}N)}\bigg\{12\bigg(G_{0}+G_{1}\big(N+8(G_{0}+G_{1}N)^{2\theta}\theta\big)\bigg)+\frac{24^{n}A(-1+n)nG_{1}\theta}{\big(G_{0}+G_{1}(N+\theta)\big)^{3}}\\
&\times\Big(-\big(G_{0}+G_{1}(N+\theta)\big)\Big)^{n}\bigg[4G_{0}^{3}+G_{1}G_{0}^{3}\Big(15+12N+48\theta-64n\theta\Big)+G_{1}^{2}G_{0}\bigg(-8+12N^{2}\\
&+\theta\Big[5+32n(1-4\theta)+84\theta\Big]+N\Big(30+32(3-4n)\theta\Big)\bigg)+G_{1}^{3}\bigg(4N^{3}+N^{2}\big(15+48\theta-64n\theta\big)\\
&-2\theta(-1+4\theta)\big(-2n-5\theta+8n\theta\big)+N\Big[-8+\theta\Big(5+32n(1-4\theta)+84\theta\Big)\Big]\bigg)\bigg]\bigg\}.
\end{split}
\end{equation}
The potential \eqref{eq42} can be rewritten in terms of the scalar field $\phi$ by considering assumptions such as $\frac{dN}{dt}=H(N)$ and $\phi=t$. In this case, we have,
\begin{equation}\label{eq44}
\begin{split}
&V(\phi)=\frac{1}{3}\Big((G_{1}+G_{0})^{1-\theta}-G_{1}(-1+\theta)\phi\Big)^{\frac{1}{-1+\theta}}\bigg\{\varepsilon+3\bigg(\bigg[(G_{1}+G_{0})^{1-\theta}-G_{1}(-1+\theta)\phi\bigg]^{\frac{1}{1-\theta}}\bigg)^{2\theta}\\
&\times\bigg(2G_{1}\theta+3\bigg[(G_{1}+G_{0})^{1-\theta}-G_{1}(-1+\theta)\phi\bigg]^{\frac{1}{1-\theta}}\bigg)-\frac{24^{n}A(-1+n)}{\bigg(G_{1}\theta+\bigg[(G_{1}+G_{0})^{1-\theta}-G_{1}(-1+\theta)\phi\bigg]^{\frac{1}{1-\theta}}\bigg)^{3}}\\
&\times\bigg(-\bigg[\Big((G_{1}+G_{0})^{1-\theta}-G_{1}(-1+\theta)\phi\Big)^{\frac{1}{1-\theta}}\bigg]^{-1+4\theta}\bigg[G_{1}\theta+\Big((G_{1}+G_{0})^{1-\theta}-G_{1}(-1+\theta)\phi\Big)^{\frac{1}{1-\theta}}\bigg]\bigg)^{n}\\
&\times\bigg[nG_{1}^{4}\theta^{2}(-1+4\theta)\Big(-\theta+n(-1+4\theta)\Big)+G_{1}^{3}\theta\bigg(n(2-5\theta)+3\theta^{2}+8n^{2}\theta(-1+4\theta)\bigg)\\
&\times\bigg(\Big[G_{1}+G_{0}^{1-\theta}-G_{1}(-1+\theta)\phi\Big]^{\frac{1}{1-\theta}}+3\Big((G_{1}+G_{0})^{1-\theta}-G_{1}(-1+\theta)\phi\Big)^{\frac{-4}{-1+\theta}}\\
&+(9+8n)G_{1}\theta\Big((G_{1}+G_{0})^{1-\theta}-G_{1}(-1-\theta)\phi\Big)^{\frac{-3}{-1+\theta}}+G_{1}^{2}\theta(-6n+(9+4n))\theta\bigg)\\
&\times\Big((G_{1}+G_{0})^{1-\theta}-G_{1}(-1+\theta)\phi\Big)^{\frac{-2}{-1+\theta}}\bigg]\bigg\}.
\end{split}
\end{equation}
The first and second potential derivatives can be obtained using equation \eqref{eq44} to call swampland dS conjectures. By placing the equations \eqref{eq44} and their derivatives in refined swampland dS conjecture, two equations of swampland dS conjectures are calculated in terms of the scalar field $\phi$. After calculating the above values, which include potential, slow-roll parameters, and two cosmological parameters $n_{s}$ and $r$, we now want to challenge a series of constraints. The primary purpose of this paper is to examine specific limitations from the mimetic f(G) gravity point of view and swampland dS conjectures. So, we first rewrite $ \phi-n_{s} $ and $ \phi-r $ according to the equations \eqref{eq40} and \eqref{eq41}, and by placing these structures in the refined swampland dS conjecture, we reconstruct $C_ {1,2} -n_ {s} $ and $ C_ {1,2 }-r$. We will study new constraints and plot figures to determine the range of these cosmological parameters and the swampland dS conjectures coefficient. Due to the length of the calculations, we determine the permissible scope of these parameters by plotting some figures. Also, according to equations \eqref{eq377}, \eqref{eq388}, \eqref{eq40} and \eqref{eq41}, we examine the restrictions of these two cosmological parameters to each other. The corresponding scalar potential has been reconstructed by analyzing the inflationary model, whereby we have assumed $f(\mathcal{G})=A\mathcal{G}^{n}\propto\mathcal{G}^{n}$ for simplicity. So, we presume the parameter $(A)$ is the positive constant of the order of one.
\begin{figure}[h!]
\begin{center}
\subfigure[]{
\includegraphics[height=6cm,width=6cm]{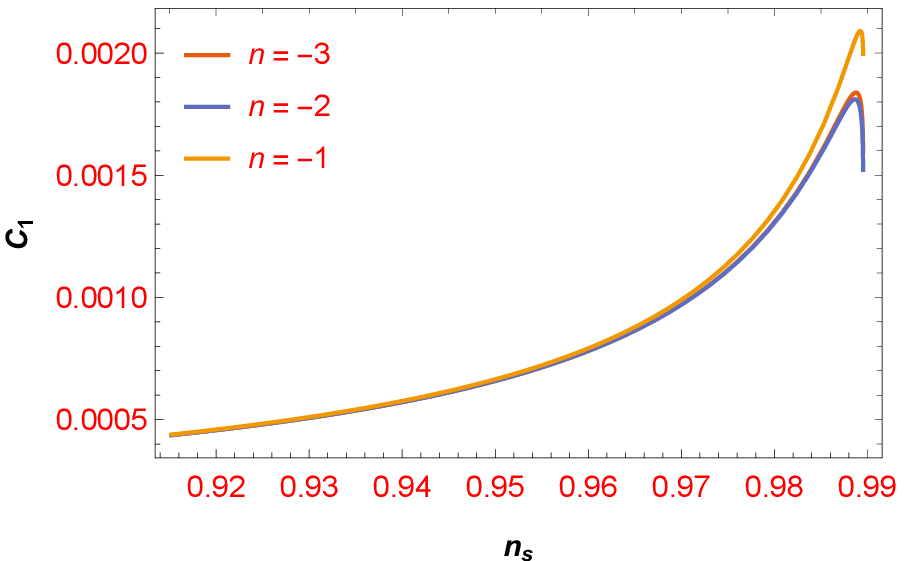}
\label{1a}}
\subfigure[]{
\includegraphics[height=6cm,width=6cm]{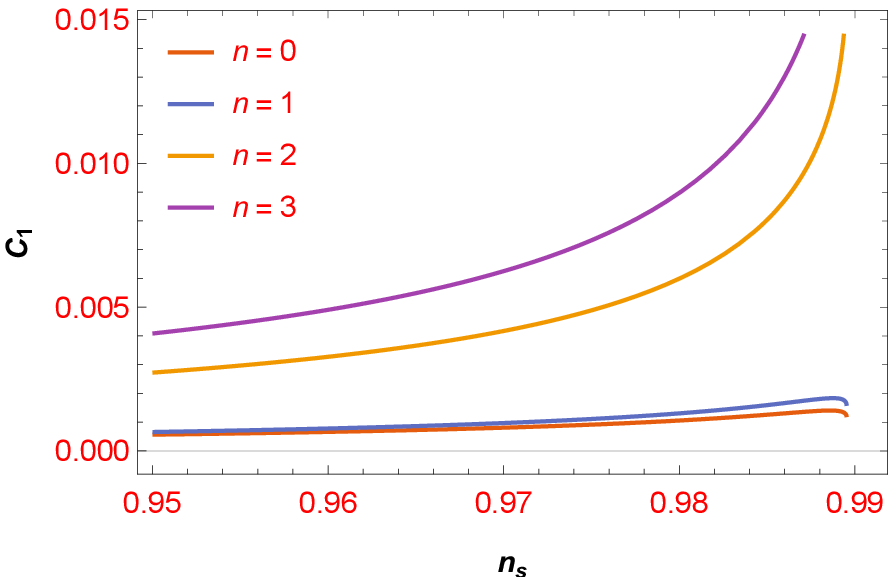}
\label{1b}}
\caption{\small{The plot of $C_{1}$ in terms of $n_{s}$ with respect to $n<0$, $n\geq0 $ and free parameter values as $(G_{0}=0.75, G_{1}=-0.00699, \theta=0.2 \hspace{0.2cm} and \hspace{0.2cm} N=60)$ }}
\label{1}
\end{center}
\end{figure}
\begin{figure}[h!]
\begin{center}
\subfigure[]{
\includegraphics[height=6cm,width=6cm]{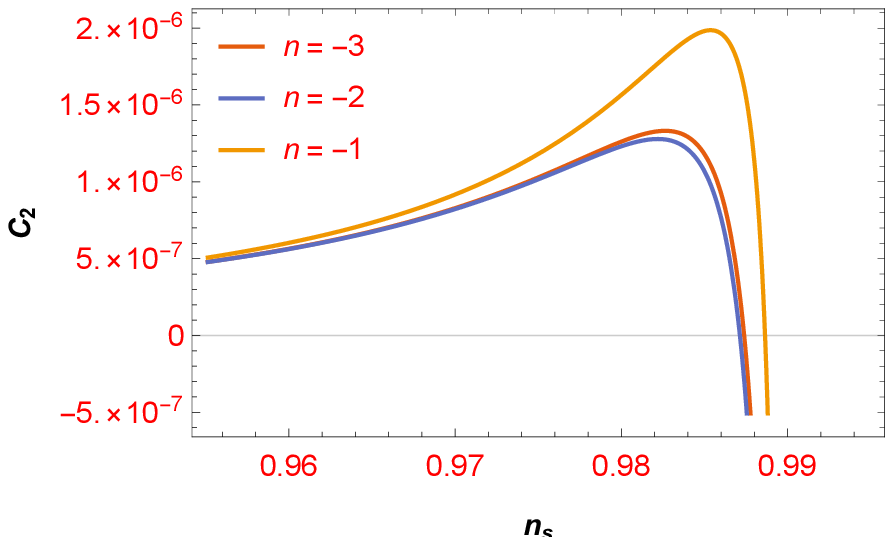}
\label{2a}}
\subfigure[]{
\includegraphics[height=6cm,width=6cm]{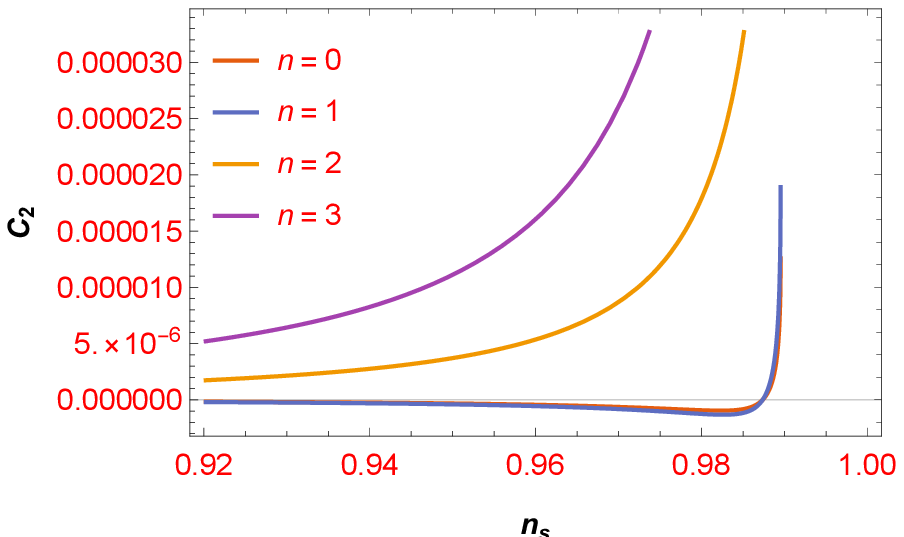}
\label{2b}}
\caption{\small{The plot of $C_{2}$ in terms of $n_{s}$ with respect to  $n<0$, $n\geq0$ and free parameter values as $(G_{0}=0.75, G_{1}=-0.00699, \theta=0.2 \hspace{0.2cm} and \hspace{0.2cm} N=60)$ }}
\label{2}
\end{center}
\end{figure}
\begin{figure}[h!]
\begin{center}
\subfigure[]{
\includegraphics[height=6.5cm,width=6.5cm]{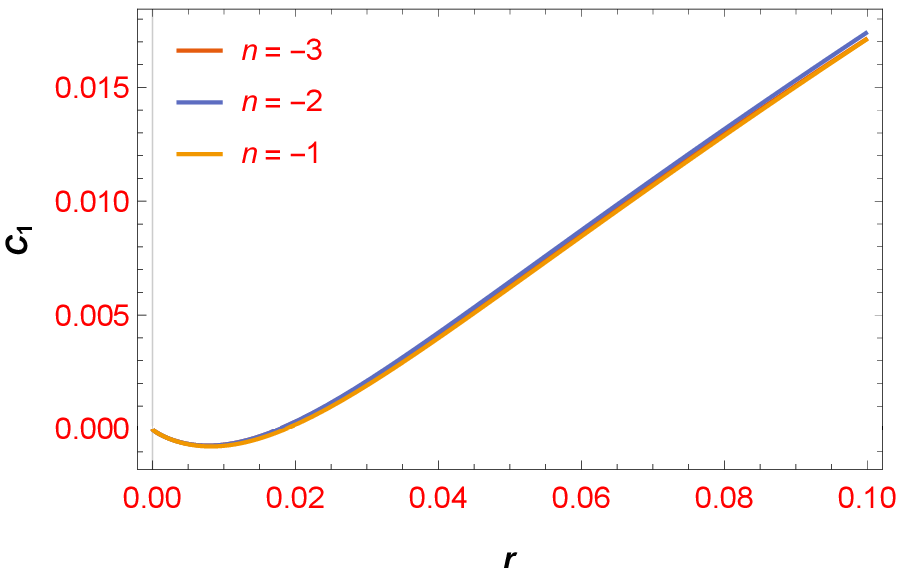}
\label{3a}}
\subfigure[]{
\includegraphics[height=6.5cm,width=6.5cm]{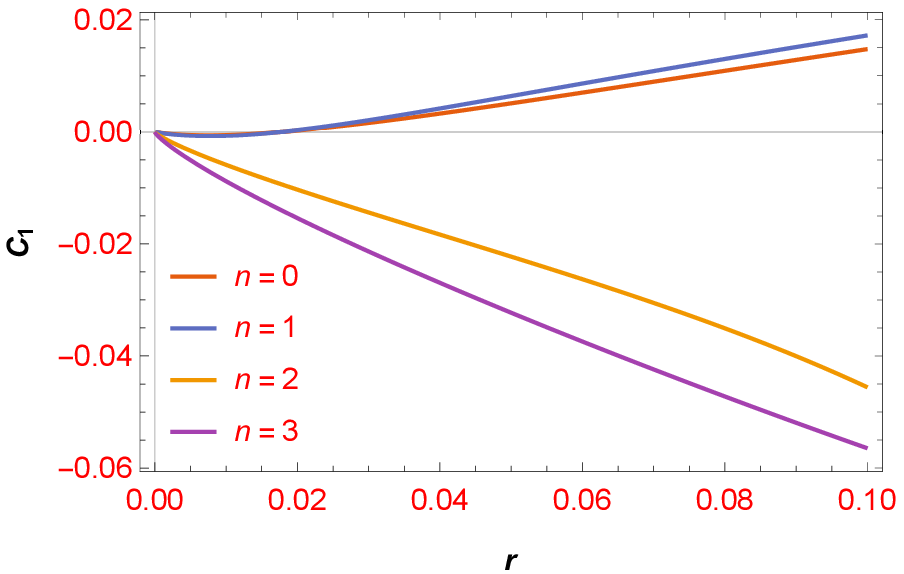}
\label{3b}}
\caption{\small{The plot of $C_{1}$ in terms of $r$ with respect to $n<0$, $n\geq0$ and free parameter values as $(G_{0}=0.75, G_{1}=-0.00699, \theta=0.2 \hspace{0.2cm} and \hspace{0.2cm} N=60)$ }}
\label{3}
\end{center}
\end{figure}
\begin{figure}[h!]
\begin{center}
\subfigure[]{
\includegraphics[height=6.5cm,width=6.5cm]{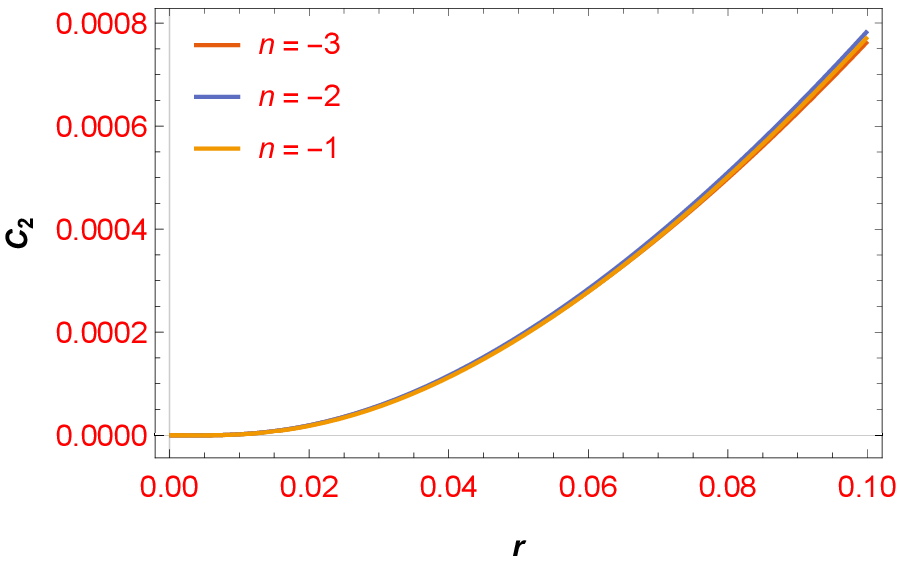}
\label{4a}}
\subfigure[]{
\includegraphics[height=6.5cm,width=6.5cm]{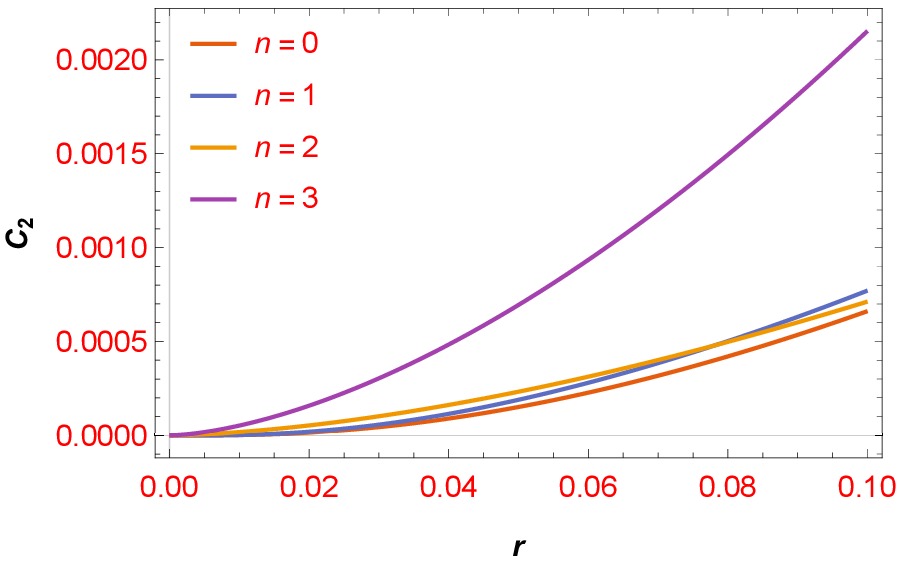}
\label{4b}}
\caption{\small{The plot of $C_{2}$ in terms of $r$ with respect to $n<0$, $n\geq0$ and free parameter values as $(G_{0}=0.75, G_{1}=-0.00699, \theta=0.2 \hspace{0.2cm} and \hspace{0.2cm} N=60)$ }}
\label{4}
\end{center}
\end{figure}
\begin{figure}[h!]
 \begin{center}
 \subfigure[]{
 \includegraphics[height=6cm,width=6cm]{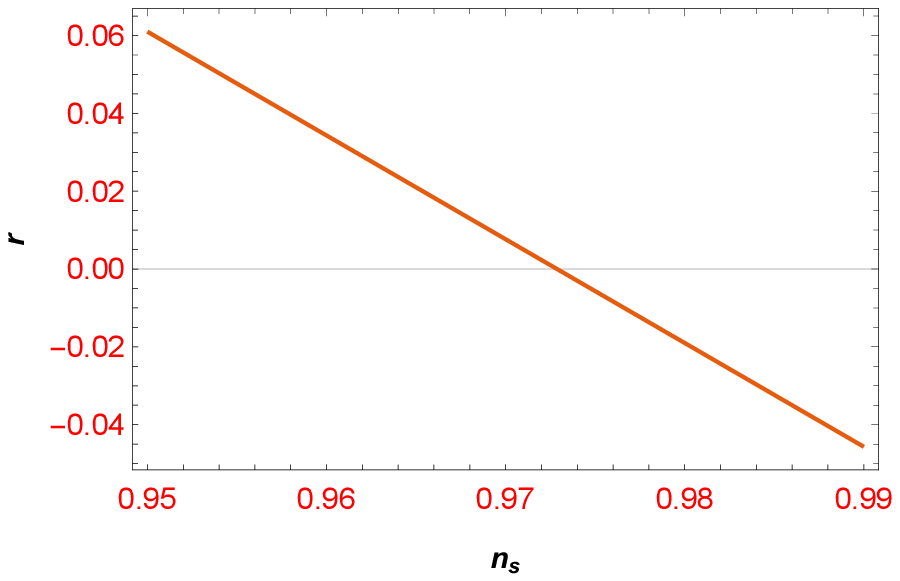}
 \label{5}}
 \caption{The plot of $r$ in term of $n_{s}$ with respect to the free parameter values as $(G_{0}=0.75, G_{1}=-0.00699, \theta=0.2 \hspace{0.2cm} and \hspace{0.2cm} N=60)$}
 \label{5}
 \end{center}
 \end{figure}
According to the above explanations, each of the figures can be analyzed. For example, figures (1) and (2) show the restrictions of the swampland dS conjecture in terms of the scalar spectral index concerning the values of free parameters that we specified for this model in the previous section and different values of n. Figure (1) clearly shows the limitations of the $C_{1}$ in the two ranges, i.e., $(-3\leq n\leq-1)$ and $(0\leq n\leq3)$. As you can see, the range of the swampland dS conjecture for $n<0$ has smaller than $n\geq 0$. Also, the $n<0$ figures have very close values that can be seen in $n \geq0$ with a slight difference. That is, we have a difference between $(n<0)$ and  $(0\leq n)$. Also, the allowable range for the scalar spectral index is shown. In figure (2), the same restrictions are applied this time for ($ C_{2} $). In this figure, even though the values for $C_{0}$ in $(n<0)$ are smaller than  $(0\leq n)$, they generally have smaller values than $C_{1}$. Also, the allowable range for this component is ($ C_{2} $), and the scalar spectral index for both ranges $(n<0)$ and  $(0\leq n)$ are well visible in the figures. However, figures (3) and (4) show the changes in the swampland dS conjecture component in terms of the tensor-to-scalar ratio for the two ranges $(n<0)$ and  $(0\leq n)$. In figure (3) indicates the limit allowed for ($ C_{1} $) in terms of (r). Also, the allowable range for the other parameter, namely  ($ C_{2} $),  for both $(n<0)$ and $n\geq0$, and for different values n, can be well defined. To determine this constraint, we also use the allowable values of the free parameters that we specified earlier as $(G_{0}, G_{1},\theta )$. Figure (5) restricts the two cosmological parameters of the scalar spectral index and the tensor-to-scalar ratio. According to the values of the free parameters in the previous section, we specify the exact values of these two cosmological parameters compatible with observable data\cite{4,5}. The important point in these studies is that the results are close to observable data. Of course, such conditions can be examined by combining them with conjectures and conditions such as constant-roll or challenging other theories. The results can be compared with the values obtained in this paper. So, one can identify the best model consistent with the observable data. These calculations can also be studied with other swampland conjectures, such as TCC. It may be possible to determine the relationship between these conjectures to advance these goals in future work.
\subsection{Further Refining Swampland dS Conjecture}
After examining the model and calculating the quantities, such as potential and expressing the coefficient of swampland dS conjectures in terms of $n_{s}$ and $r$ to determine the allowable range for each of these cosmological parameters by plotting some figures; here, we want to challenge the newer conjecture recently introduced in the literature by David Andriot and Christoph Roupec \cite{a}, which is called further refining swampland dS conjecture. Further refined dS conjecture is considered for effective low-energy theories of quantum gravity, provided that the potential is larger than zero at any point in the field space \cite{67,68,a,68ab}, which is as follows:
\begin{equation}\label{eq45}
\left(M_{pl}\frac{|V'| }{V}\right)^q-a M_{pl}^2 \frac{|V''|}{V}\geq b  \hspace{1cm}   with \hspace{1cm}  a,b>0\hspace{0.5cm},q>2, \hspace{0.5cm} a+b=1.
\end{equation}
The above relation is a combination of the first and second derivatives of the scalar potential. We can  rewrite this conjecture in terms of slow-roll parameters as follows,
\begin{equation}\label{eq46}
(2\epsilon)^{\frac{q}{2}}-a\eta \geq b.
\end{equation}
The above relationship has been examined in  \cite{67} and \cite{68} for the Higgs inflation and  Inflationary models in general scalar-tensor theory, respectively. We define the two parameters $F_1$ and $F_2$, which are other forms of slow-roll parameters and can be helpful, as follows,
\begin{equation}\label{eq47}
F_1=\frac{|V'(\phi)|}{V(\phi)}=\sqrt{2\epsilon} \hspace{1.5cm}   F_2=\frac{V''(\phi)}{V(\phi)}=\eta
\end{equation}
In this case, we rewrite equation \eqref{eq46} in terms of $F$,
\begin{equation}\label{eq48}
(F_1)^{\frac{q}{2}}-a F_2 \geq 1-a
\end{equation}
Also, with a simple calculation, rewrite $F_1$ and $F_2$ in terms of  $r$ and $n_s$, which are e spectrum index of the primordial curvature power spectrum and d tensor-to-scalar ratio $r$, respectively, which are given by,
\begin{equation}\label{eq49}
F_1=\sqrt{2\epsilon}=\sqrt{\frac{r}{8}} \hspace{1.5cm}   F_2=\eta=\frac{1}{2}(n_s-1+\frac{3r}{8}).
\end{equation}
Now, we examine the mimetic $f(G)$ gravity inflation model in terms of the further refining swampland dS conjecture of whether it met.
The scalar spectral index and the tensor-to-scalar ratio are obtained $n_s=0.964569$ and $r=0.06836$ for mimetic $f(G)$ gravity inflation model.
By putting these values in Equation \eqref{eq49}, we get,
\begin{equation}\label{eq50}
F_1=\sqrt{2\epsilon}=\sqrt{\frac{r}{8}}=0.09243, \hspace{1.5cm}   F_2=\eta=\frac{1}{2}(n_s-1+\frac{3r}{8})=-0.00489.
\end{equation}
Considering the refined swampland dS conjecture, we find:
\begin{equation}\label{eq51}
C_{1}\leq 0.09243, \hspace{1.5cm}   C_2 \leq 0.00489,
\end{equation}
$C_{1}$ and $C_2$ are not  $\mathcal{O}(1)$; as a result, refining swampland ds conjecture is not satisfied for this inflationary model. We now examine it with further refining swampland dS conjecture. By placing equation \eqref{eq50} inside equation \eqref{eq48}, we get,
\begin{equation}\label{eq52}
(0.09243)^q+ 0.00489 a \geq 1-a \hspace{1.5cm} or \hspace{1.5cm} (0.09243)^q \geq 1- 1.00489 a
\end{equation}
We can find a relation to satisfying the recent condition,
\begin{equation}\label{eq53}
\frac{1}{1.00489}(1-(0.09243)^q)\leq a <1 ,  \hspace{1cm} q>2.
\end{equation}
In this case,  the further refining swampland dS conjecture is satisfied. According to relation \eqref{eq52}, when $a < \frac{1}{1.00489}=0.99513$, we can always find a $q$ whose value is larger than 2. For example, for $q=2.4$ with respect to relation \eqref{eq53}, we find $0.99185\leq a < 1$, which we can choose $a=0.99235$ according to the condition $a < 0.99513$. Also, we know $b=1-a$, so we will have $1-0.99235=0.00765 > 0$.
\section{Conclusion}
Mimetic gravity analysis has been studied as a theory in various types of general relativity extensions, such as mimetic f(R) gravity,  mimetic f(R, T) gravity, mimetic f(R, G) gravity, etc., in the literature. This paper presented a set of equations arising from mimetic conditions. We studied cosmic inflation with mimetic f(G) gravity and swampland dS conjectures. We analyzed and evaluated these results thoroughly. Therefore, we first introduced the mimetic f(G) gravity thoroughly and calculated some cosmological parameters such as the scalar spectral index ($n_{s}$) and the tensor-to-scalar ratio ($r)$, and the slow-roll parameters. Then we investigated the potential according to the mimetic f(G) gravity. Then we challenged the swampland dS conjectures with this condition. By limiting the coefficient of swampland dS conjectures in terms of $n_{s}$ and $r$, we plotted some figures and determined the allowable range for these cosmological parameters. Finally, we compared these results with observable data such as Planck and BICEP2/Keck array data\cite{4,5}.  Also, we showed $C_{1}$ and $C_2$ were not $\mathcal{O}(1)$; as a result, refining swampland dS conjecture was not satisfied for this inflationary model. Then we examined it by further refining the swampland dS conjecture; by adjusting their free parameters, we discussed the compatibility of the mentioned conjecture with the inflation model and ultimately stated the results. The studies in this paper can be thoroughly reviewed for all other theories of mimetic gravity, such as f(R), f(R, T), f(R, G) gravity, etc. So The results can be compared with the observational data and other theories. This study can lead to exciting results. Also, this model or other inflationary models(mimetic gravity) can be examined from the point of view of other swampland conjectures, such as distant conjectures and TCC. The results can be analyzed entirely. It is also possible to obtain interesting results by combining other conditions, such as the constant-roll condition with mimetic gravity and swampland conjectures, all of which are points that can be considered for future work.\\

\section{Acknowledgments}
This work is dedicated to the memory of Prof. Farhad Darabi.\\

\end{document}